\documentclass[twocolumn]{aastex63}
\usepackage{graphics,epsf}
\usepackage{amsmath}                
\usepackage{amsfonts}               
\usepackage{amssymb}                
\usepackage{epsfig}                 
\usepackage{appendix}
\usepackage{graphicx}
\usepackage{float}
\usepackage{color}
\usepackage{multirow}
\usepackage{colortbl}
\usepackage[para,online,flushleft]{threeparttable}

\hypersetup{citecolor=blue, 
            linkcolor=red, 
            menucolor=blue, 
            urlcolor=blue}  

\newcommand{\kms}{{~\rm km\; s^{-1}}}

\newcommand{\cm}{{~\rm cm}}
\newcommand{\m}{{~\rm m}}
\newcommand{\km}{{~\rm km}}
\newcommand{\s}{{~\rm s}}
\newcommand{\h}{{~\rm h}}

\newcommand{\g}{{~\rm g}}
\newcommand{\G}{{~\rm G}}
\newcommand{\K}{{~\rm K}}
\newcommand{\erg}{{~\rm erg}}
\newcommand{\yr}{{~\rm yr}}

\newcommand{\AU}{{~\rm AU}}

\newcommand{\days}{{~\rm days}}

\begin{document}

\title{Rapid decline in the lightcurves of luminous supernovae by jet-driven bipolar explosions}


\author[0000-0001-7233-6871]{Muhammad Akashi}
\affiliation{Department of Physics, Technion, Haifa, 3200003, Israel; akashi@physics.technion.ac.il; soker@physics.technion.ac.il}
\affiliation{Kinneret College on the Sea of Galilee, Samakh 15132, Israel}

\author[0000-0002-1361-9115]{Amir Michaelis}
\affiliation{Department of Physics, Ariel University, P.O. Box 3, Ariel 4070000, Israel; amirmi@ariel.ac.il}

\author[0000-0003-0375-8987]{Noam Soker}
\affiliation{Department of Physics, Technion, Haifa, 3200003, Israel; akashi@physics.technion.ac.il; soker@physics.technion.ac.il}

\begin{abstract}
We calculate the lightcurves of jet-driven bipolar core collapse supernova (CCSN) explosions into a bipolar circumstellar mater (CSM) and show that an equatorial observer finds the lightcurves to possess a rapid, and even an abrupt, drop.  The scenario that might lead to such an explosion morphology is a common envelope evolution (CEE) where shortly before the CCSN explosion the red supergiant  progenitor interacts with a more compact companion that spirals-in and spins-up the core. The companion can be a main sequence star, a neutron star, or a black hole. The binary interaction ejects a shell through an intensive wind and the CEE ejects a denser gas in the equatorial plane. We assume that the companion accretes mass and launches jets. We conduct three-dimensional (3D) hydrodynamical simulations where we launch weak jets, the shaping jets, into the dense shell and show that the interaction forms a bipolar CSM. As a result of the rapid pre-collapse core rotation jets drive the CCSN explosion. We simulate the interaction of the jets with the bipolar CSM and use a simple scheme to calculate the lightcurves. We show that the abrupt drop in the lightcurve of an observer not too close to the polar directions can account for the lightcurve of the hydrogen poor luminous supernova (LSN) SN 2018don. 
 \end{abstract}


\section{Introduction} 
\label{sec:intro}

Many studies have reached the conclusion that superluminous supernovae (SLSNe), i.e., core collapse supernovae (CCSNe) with peak r-band magnitude of $M_{\rm r} < -20$, and luminous supernovae (LSNe) that are CCSNe with $M_{\rm r} =-19$ to $-20$ (see \citealt{Gomezetal2022} for definitions), contain an extra energy source or another ingredient in addition to the explosion itself. We can distinguish between two classes of explanations. 

In the first class of explanations the explosion itself is the delayed neutrino mechanism. It seems that the heavily studied delayed neutrino mechanism (\citealt{BetheWilson1985, Hegeretal2003, Janka2012, Nordhausetal2012, Bruennetal2016, Jankaetal2016R, OConnorCouch2018, Mulleretal2019Jittering, BurrowsVartanyan2021, Fujibayashietal2021, Bocciolietal2022, Nakamuraetal2022}; for a recent brief optimistic review of this mechanism see \citealt{Mezzacappa2022}) cannot reach CCSN explosion energies of $E_{\rm SN} \ga 2 \times 10^{51} \erg$ (e.g., \citealt{Fryer2006, Fryeretal2012, Papishetal2015a, Sukhboldetal2016, SukhboldWoosley2016, Gogilashvilietal2021}). Many studies of SLSNe and LSNe that take the explosion itself to be by the delayed neutrino mechanism attribute the extra energy of LSNe and SLSNe to a magnetar, i.e., a rapidly rotating magnetized neutron star (e.g., \citealt{Greineretal2015, Metzgeretal2015, Yuetal2017, Marguttietal2018, Nicholletal2017b}). 

According to the second class of explanations jets explode most (or even all) CCSNe (for a review see \citealt{Soker2022Rev}). Even if the pre-collapse core is slowly rotating, accretion of material with stochastic angular momentum onto the newly born neutron star leads to the launching of jittering jets (jets with stochastic variation of directions), i.e., the jittering jets explosion mechanism (\citealt{Soker2010, PapishSoker2014Planar, GilkisSoker2015,  Quataertetal2019, Soker2019RAA, Soker2020RAA, ShishkinSoker2021, AntoniQuataert2022, Soker2022a, Soker2022Boosting}). Therefore, according to this class of models jets also explode all LSNe and SLSNe, as well as the CCSNe that accompany gamma ray bursts as \cite{Izzoetal2019} also claim from their observation of a gamma ray burst {{{{ and in line with earlier studies of jet-driven gamma ray bursts (e.g., \citealt{ShavivDar1995, MacFadyenWoosley1999}). }}}}  
Further motivation to consider jet-driven explosions is the problems of magnetar modelling of some SLSNe, e.g., \cite{SokerGilkis2017a} who analysed the magnetar modelling by \cite{Nicholletal2017b}, and LSNe, e.g., \cite{Soker2022LSNe} who analysed the magnetar modelling by \cite{Gomezetal2022}.
Moreover, some studies find that the formation of an energetic magnetar most likely leads to the launching of jets during the explosion and in some cases also after the explosion, and that these jets might carry more energy that the magnetar \citep{Soker2016Magnetar, Soker2017Magnetar, SokerGilkis2017a}; for some possible energy sources of SLSNe see, e.g., \citealt{WangWangDai2019RAA}.) 
 
According to the jittering jets explosion mechanism the jets that explode LSNe and SLSNe keep more or less a fixed axis. Namely, the jittering is  small. This requires the pre-collapse core to have fast rotation. For example, a binary companion that spirals-in inside the red supergiant progenitor to a very small orbital separation both spins-up the core and removes the hydrogen-rich envelope to lead eventually to striped-envelope CCSN (CCSN-I), i.e., a type Ib or a type Ic CCSN. The fixed axis makes the efficiency of the jet feedback mechanism (for a review see \citealt{Soker2016Rev}) inefficient, namely, the explosion energy will be many times the binding energy of the exploding star. Therefore, the  explosion is very energetic \citep{Gilkisetal2016Super, Soker2017RAA}. 

Note that unlike many studies that require the pre-collapse core to have very fast rotation so its specific angular momentum by itself allows the formation of accretion disk around the newly born neutron star or black hole (e.g., \citealt{BisnovatyiKogan1971, DuncanThompson1992, Khokhlovetal1999, MacFadyenWoosley1999, Aloyetal2000, MacFadyenetal2001, Wheeleretal2002, Akiyamaetal2003, Burrowsetal2007, Wheeleretal2008, Maedaetal2012, Winteleretal2012, LopezCamaraetal2013, Mostaetal2014, BrombergTchekhovskoy2016,  Nishimuraetal2017, Sobacchietal2017, Grimmettetal2021, Gottliebetal2022, Perleyetal2022}), 
in the jittering jets explosion mechanism convective fluctuations in the pre-collapse core can supply additional stochastic angular momentum and therefore the core might have a slower rotation  (e.g., \citealt{GilkisSoker2014, GilkisSoker2016, ShishkinSoker2022}). 

\cite{KaplanSoker2020b} proposed a toy model to explain the puzzling lightcurve of the LSNe SN~2018don. SN~2018don is a hydrogen-poor LSN that has a sharp drop in its lightcurve \citep{Lunnanetal2020}. It has a long rise time to maximum light relative to other LSNe \citep{Gomezetal2022}.
\cite{KaplanSoker2020b} built simple bipolar ejecta models for jet-driven CCSNe and showed that for an observer in the equatorial plane of the bipolar morphology the lightcurve has a rapid luminosity decline, and even an abrupt drop. In some cases there is no abrupt drop, but rather a `knee' in the lightcurve where a rapid drop changes to a moderate drop in luminosity. \cite{Soker2022LSNe} proposes that such a bipolar explosion might explain the `knee' in the lightcurve of the SLSN SN~2020wnt (for its properties see \citealt{Tinyanontetal2021, Gutierrezetal2022, Tinyanontetal2022}).

\cite{KaplanSoker2020b} did not simulate the bipolar jet-driven explosion, but rather simply assumed that the ejecta has an axisymmetrical structure composed of an equatorial ejecta and faster polar ejecta, each which is a part of a sphere (different spheres), and that it has a uniform effective temperature. In this study we conduct hydrodynamical simulations to simulate the bipolar explosion (section \ref{sec:Bipolar}), and then estimate the lightcurve, assuming a uniform effective temperature (because we do not calculate radiative transfer) but only if the photosphere is hot (section \ref{sec:Lightcurve}). 
We summarize our study that strengthens the claim that jets power (at least some) LSNe and SLSNe, and maybe even most CCSNe (section \ref{sec:Summary}). 

\section{The bipolar jet-driven explosion} 
\label{sec:Bipolar}
In this section we describe numerical simulations of a jet-driven bipolar CCSN explosion into several different compact bipolar circumstellar matter (CSM) morphologies.  

\subsection{The numerical code} 
\label{subsec:Numerical}

We use version 4.6.2 of the adaptive-mesh refinement (AMR) hydrodynamical FLASH code \citep{Fryxell2000} in three dimension (3D). We turn off the radiative cooling since the regions we simulate are optically thick. 
We include radiation pressure, electrons pressure, and ions pressure as we assume an adiabatic index of $\gamma=5/3$ and $\mu = 0.5$, as we assume a fully ionised pure hydrogen. 
We set outflow conditions on all the boundary surfaces of the 3D grid. We use resolution with 7 refinement levels, i.e., the ratio of the length of the sides of the grid to the shortest cell in the grid is $2^{9}$. The finest cell in the grid has a side of $1.05 AU$. We simulate the whole space with a total size of the Cartesian  numerical grid of $(534 AU)^3$, i.e., $(L_x,L_y,L_z) = \pm 267 AU$.
We simulate the entire space, rather than only one half of the symmetry plane $z=0$, to avoid numerical artifacts at the symmetry plane had we simulated only one half of the interaction region. We will present the flow structure in the  entire computational grid to emphasize the bipolar structure that the jets shape and the flow in the equatorial plane.

\subsection{Building the bipolar explosion} 
\label{subsec:Explosion}

We build the bipolar explosion by three numerical phases. These might result from a common envelope evolution (CEE) or a grazing envelope evolution (GEE).
{{{{ We base our assumption of a bipolar CSM on the scenario that we study here that involves a CEE, and on the bipolar structure of many planetary nebulae that are descendant of CEE (e.g., \citealt{Corradietal2014} for observations, \citealt{Zouetal2020} for a theoretical study, and \citealt{DeMarcoIzzard} for a review). }}}} 
However, for the present study the only important phase is the general bipolar explosion with dense equatorial CSM gas and fast bipolar outflows. 

\subsubsection{The bipolar CSM} 
\label{subsubsec:Wind}

We build a bipolar CSM into which the jet-driven explosion occurs. The manner by which we built the CSM is not so important. We expect a CEE or a GEE to form such a bipolar CSM. The explosion then takes place into the bipolar CSM.
Namely, the CSM is formed and shaped, for example, by a precursor where the CCSN progenitor expands few years before the explosion and engulfs a companion that launches jets (e.g., \citealt{McleySoker2014, DanieliSoker2019, Soker2022Rev}).
Core activity of excitation of waves (e.g., \citealt{QuataertShiode2012, ShiodeQuataert2014, Fuller2017, FullerRo2018, WuFuller2022}) or magnetic activity (e.g., \citealt{SokerGilkis2017b}) causes the envelope expansion years to months before explosion.

We set the initial (at $t=0$) grid to have an expanding dense shell of mass $M_{\rm wind}$ from the center to an outer radius of $R_{\rm sh}=10^{15} \cm = 67 \AU$. The shell was formed by an intensive wind due to the onset of a CEE/GEE, i.e., the initial phase of a CEE/GEE that lasts for a few dynamical times on the surface of the red supergiant. {{{{ For the wind velocity of $v_{\rm wind} =100 \km \s^{-1}$ that we take here, the binary interaction has been blowing this wind for a period of about three years. This is about few times the dynamical time on the surface of a red supergiant star. }}}}
We present the values of this mass in the second column of Table \ref{Table:cases}. In the third column we list the velocity of the wind for each case $v_{\rm wind}$, in the forth column the corresponding mass loss rate of the wind, and in the fifth column we list the total kinetic energy in this wind. The duration of the spherical wind is $3.2(v_{\rm wind}/100 \km \s^{-1})^{-1} \yr$. The density in this shell goes as $\rho_{\rm wind} \propto r^{-2}$. 
In the region $r>67 \AU$ we have at $t=0$ a low density expanding gas with the same velocity as the wind. The spherical wind that composes the initial condition at $t=0$ is the first phase of our numerical study. 
{{{{ We assume that pre-explosion evolution processes of shocks and radiation from the CEE heat the CSM to a typical temperature of $10000\K$, as in the warm ISM and in planetary nebulae. As well, the optically thick shell implies a slow radiative cooling as photon diffusion time is long relative to the flow time. We therefore }}}} set the initial temperature of the simulation box and the jets to $10000\K$. {{{{ The jets are highly supersonic such that their initial temperature does not influence the results. }}}}
\begin{table*}
\centering
\begin{tabular}{|c|c|c|c|c|c|c|}
\hline
Sim. & $M_{\rm wind}$ & $v_{\rm wind}$ & $\dot{M}_{\rm wind}$ & $E_{\rm wind}$ & $M_{\rm jets}$ & $E_{\rm jets}$  \\ 
           & $M_\odot$ & $\kms$ & $M_\odot \yr^{-1}$ & $\erg$ & $M_\odot$ & $\erg$\\ 
 \hline 
EXP1       & 0.25 & 100 & 0.08 &$2.5 \times 10^{47}$ & $3.3 \times 10^{-5}$ & $8.2 \times 10^{47}$ \\ \hline
EXP2      & 2.5 & 100 & 0.8 &$2.5 \times 10^{48}$  & $3.3 \times 10^{-4}$ & $8.2 \times 10^{48}$\\ \hline
EXP3       & 0.75  &57.7 & 0.24 &$2.5 \times 10^{47}$ & $ 10^{-4}$ & $8.2 \times 10^{47}$\\ \hline
EXP4      & 0.75  &40.8 &0.24 &$1.25 \times 10^{47}$ & $10^{-4}$ & $4.1 \times 10^{47}$\\ \hline

\end{tabular}
\caption{Summary of the CSM properties for the four simulations that we present in the paper. The columns list, from left to right and for each simulation, the name of the simulation, total mass of the spherical wind, the velocity of the wind, the mass loss rate of the wind and the total kinetic energy of this wind.
We start the simulations at $t=0$ by launching two opposite jets into the wind (shell) for 60 days. We list the total mass of these jets that shape the bipolar CSM just before the explosion and their total kinetic energy in the sixth and seventh column, respectively. In all simulations the jet-driven explosion energy that we launch later is $E_{\rm SN} = 2.5\times 10^{51} \erg$, and the explosion duration, basically the exploding-jet activity duration, is $3 \days$. }
\label{Table:cases}
\end{table*}

At $t=0$ we launch two opposite jets to form a bipolar CSM. The shaping of the CSM by these  jets is the second phase of our numerical study. These jets are active for 60 days in all cases.
To launch the jets we set a fast  outflow, velocity of $V_{\rm j} =  5 \times 10^4 \km \s^{-1}$ inside two opposite cones, each cone with a length of $r=7 \AU$ and a half-opening angle of $\alpha = 40^{\circ}$. The density of the jets inside the cones is $\rho_j (r) = \dot{M}_{\rm jets} / [4\pi r^{2} v_{\rm j} (1-\cos \alpha)]$, where  $\dot{M}_{\rm jets}=M_{\rm jets}/60~{\rm day}$ is the mass loss rate into the jets and $M_{\rm jets}$ is the total mass the two shaping jets carry.  
In the sixth and seventh column of Table \ref{Table:cases} we list the combined mass and energy of these jets, respectively, for the four simulations.  

During the entire simulation we inject a dense equatorial outflow due to the binary interaction. This equatorial outflow has a half opening angle of $30^{\circ}$ from the equatorial plane, a mass loss rate of $1 M_{\odot} \yr^{-1}$, and a velocity of $100 \kms$. During the entire simulation (including the later explosion phase) we inject a mass of $\approx 0.4M_\odot$ in the equatorial outflow. {{{{ The motivation to include the dense equatorial outflow comes from the presence of tori (equatorial rings) in many bipolar planetary nebulae (e.g., \citealt{DeMarcoIzzard} for a review), and the dense ring alongside the bipolar (outer) rings of SN~1987A. In any case, the bubbles that the shaping jets inflate in the wind compress gas towards the equatorial plane, such that a dense equatorial outflow is formed by the jet-inflated bubbles in the wind (e.g., \citealt{Akashietal2015}). Therefore, our qualitative results hold even without a pre-set equatorial outflow (as we wrote, we expect such an equatorial outflow to take place in many cases). }}}}

\begin{figure} 
	\centering
	\includegraphics[trim=0.0cm 3.2cm 0.0cm 4.4cm ,clip, scale=0.55]{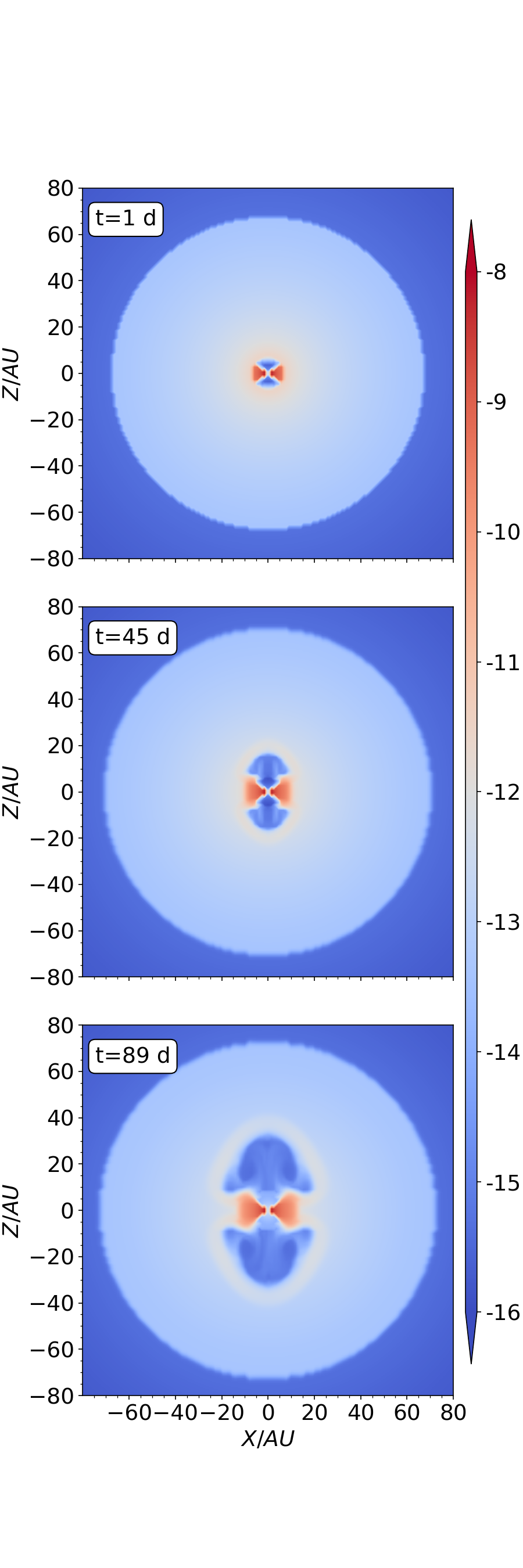}
	\caption{The density maps in the meridional plane at three times (as indicated on the panels) during the formation of the bipolar CSM into which we later explode the star. The shaping-jets are active from $t=0$ for 60 days. The logarithm of the density in units of $\rm g ~cm^{-3}$ is according to the color-bar.}
	\label{Fig:time_zero}
\end{figure}

In Fig. \ref{Fig:time_zero} we present the formation of the bipolar CSM for simulation EXP1. All panels show the density in the meridional plane, i.e., a plane along the jets' axis. The shaping-jets are active for 60 days starting at $t=0$. In the upper, middle, and bottom panels we present the density maps at $t=1$~day, $t=45$~day, and $t=89$~day, respectively.  

\subsubsection{Jet-driven explosion} 
\label{subsubsec:ExplosionJets}

The third numerical phase is a second, much more energetic, jet-launching episode. This jet-launching phase is the jet-driven CCSN explosion, as we assume that jets power this CCSN (see section \ref{sec:intro}). We assume that the newly-born neutron star accretes fallback material and launches the jets for three days starting 30 days after we turned off the jets that shaped the CSM (section \ref{subsubsec:Wind}). Namely, we launch the exploding jets during the time period $90 \days <t<93 \days$.
We launch two opposite jets into two opposite cones each cone with a length of $r=7 \AU$ and a half-opening angle of $\alpha = 40^{\circ}$. During these three days we linearly increase the jets' velocity from zero to $100,000 \kms$,
$v_{\rm ej}(t)=100,000 [t({\rm day})-90~{\rm day}]/3 \kms$, and we linearly decrease the mass loss rate into the exploding jets from $10 M_{\odot} \yr^{-1}$ to zero. 
During the explosion we continue to inject the dense equatorial outflow of a half opening angle of $30^{\circ}$ from the equatorial plane, a mass loss rate of $1 M_{\odot} \yr^{-1}$, and a velocity of $100 \kms$. 

We turn now to analyse the flow properties of the ejecta-CSM interaction, postponing the calculation of the lightcurve to section \ref{sec:Lightcurve}. 
  
\subsection{The explosion flow structure} 
\label{subsec:ExplosionStructure}
 
We describe the flow that results from the interaction of the jet-driven ejecta (section \ref{subsubsec:ExplosionJets}) with the bipolar CSM (section \ref{subsubsec:Wind}). We recall that the exploding jets are active for three days from $t=90~{\rm day}$ to $t=93~{\rm day}$. The jets' symmetry axis is along the $z$ axis through the origin.

\begin{figure}[ht!]
	\centering
	\includegraphics[trim=0.1cm 0.8cm 0.0cm 1.0cm ,clip, scale=0.44]{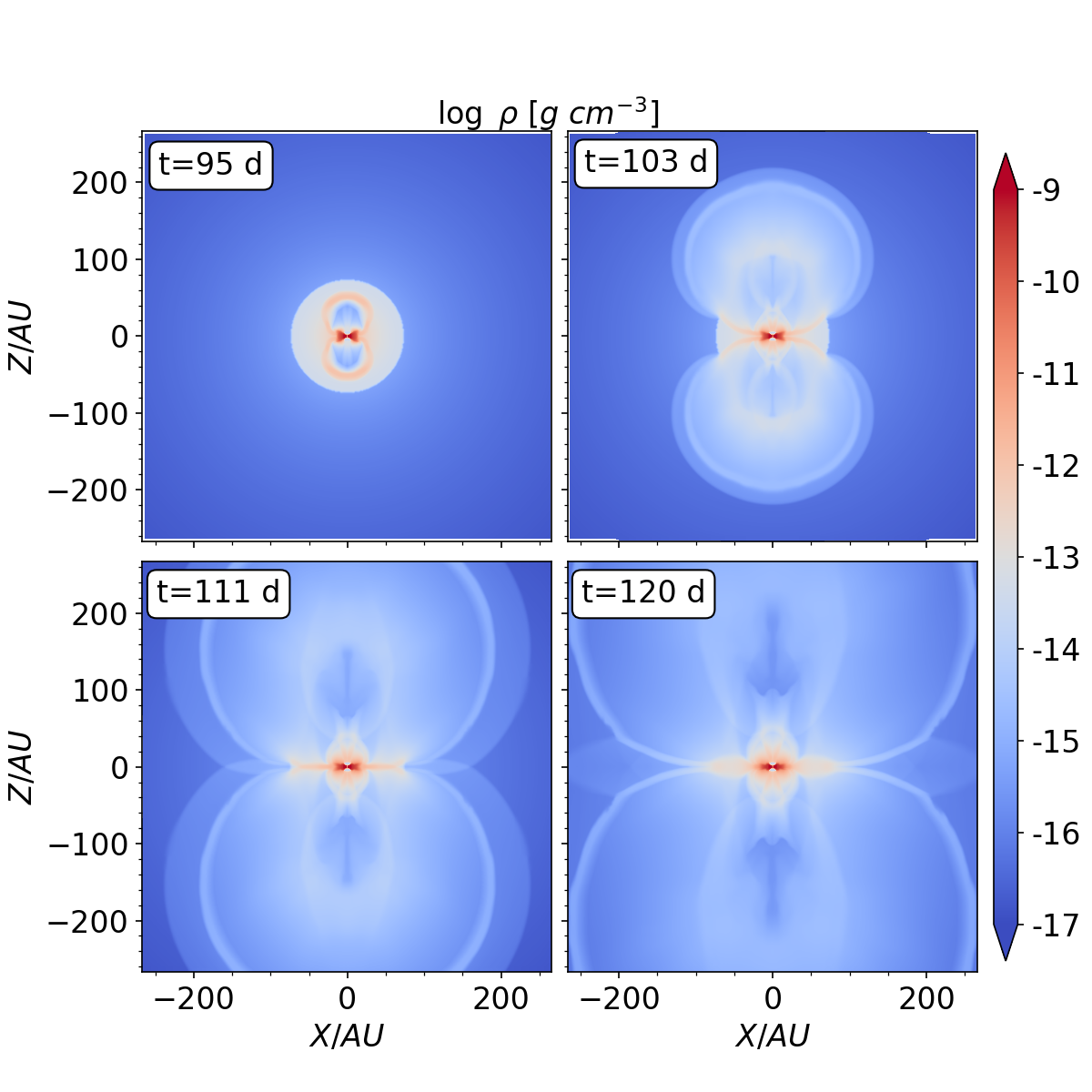}
	\caption{Density maps in the meridional plane at four times all after explosion, as indicated on the panels. Time is measured from the onset of the jets that shape the CSM. The exploding jets are active from $t=90~{\rm day}$ to $t=93~{\rm day}$. The logarithm of the density in units of $\rm g ~cm^{-3}$ is according to the color-bar. }
	\label{Fig:dens}
\end{figure}

\begin{figure}[ht!]
	\centering
	\includegraphics[trim=0.1cm 0.8cm 0.0cm 1.0cm ,clip, scale=0.44]{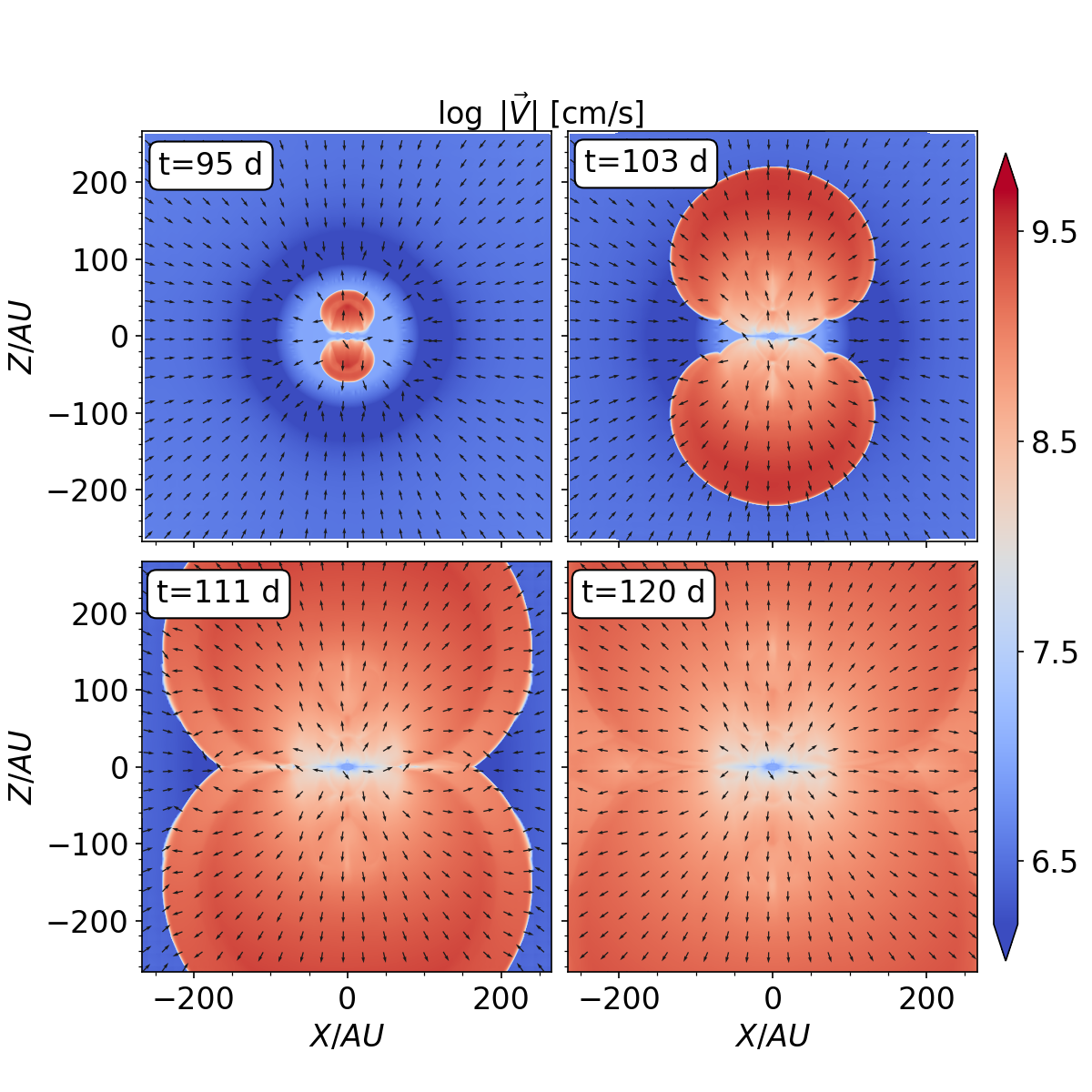}
	\caption{Similar to Fig. \ref{Fig:dens} but showing the velocity maps. The direction is according the arrows and the logarithm of the velocity magnitude in units of $\rm cm ~s^{-1}$ is according to the color-bar.  }
	\label{Fig:vel}
\end{figure}

\begin{figure}[ht!]
	\centering
	\includegraphics[trim=0.1cm 0.8cm 0.0cm 1.0cm ,clip, scale=0.44]{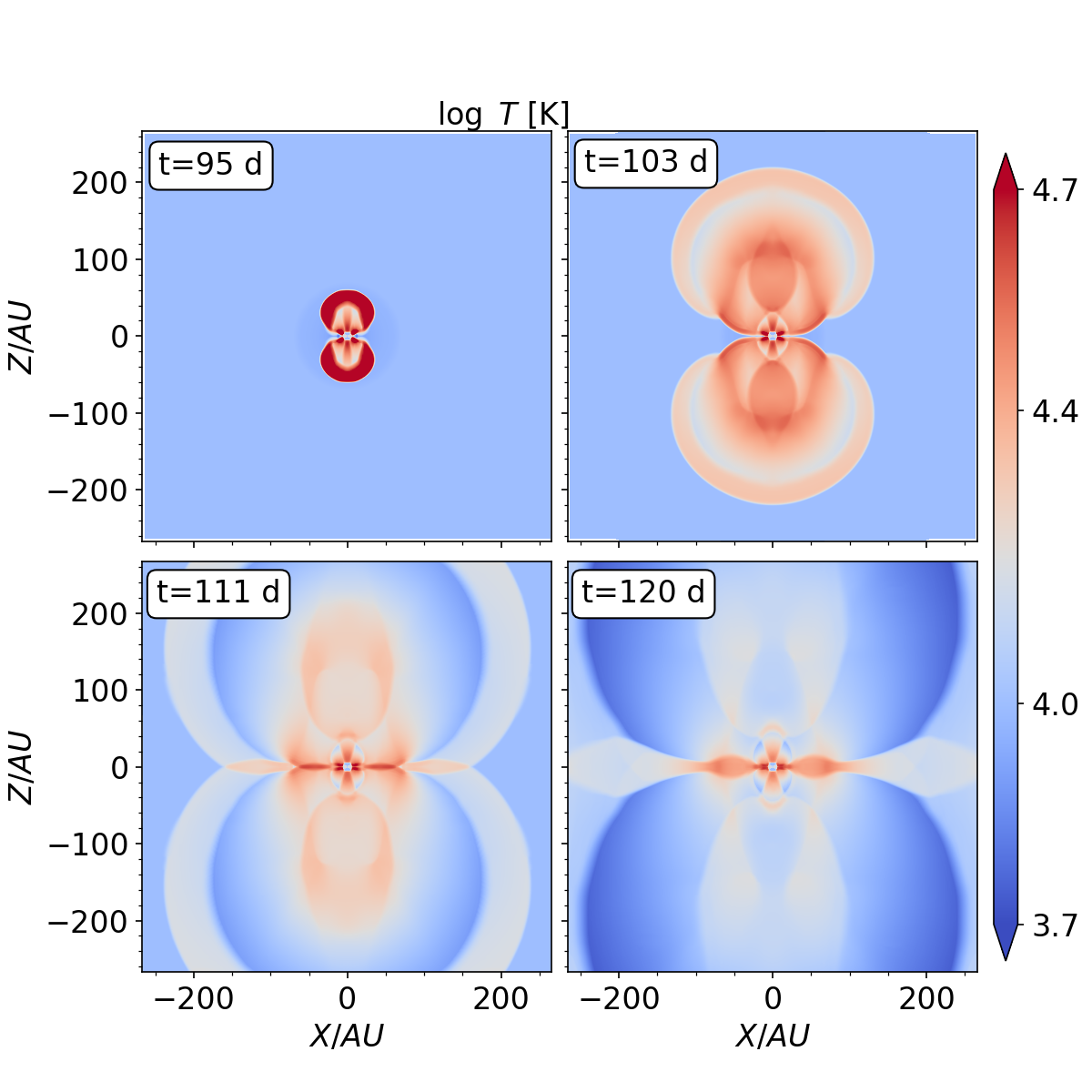}
	\caption{Similar to Fig. \ref{Fig:dens} but showing the temperature maps. The color-bar gives the logarithm of the temperature in K. }
	\label{Fig:temp}
\end{figure}

In Figs. \ref{Fig:dens}, \ref{Fig:vel} and \ref{Fig:temp} we present the density, velocity, and temperature maps, respectively, in the meridional plane at four times as the insets indicate in the panels.  

As expected, the jet-driven explosion into a bipolar CSM forms a highly non-spherical bipolar explosion. The faster expansion along and near the polar directions has implications to the lightcurve (e.g., \citealt{KaplanSoker2020b}). The first one is that the inflated polar lobes imply that at early times an equatorial observer sees a larger photosphere than a polar observer. Later, because of the faster expansion of the lobes the photosphere of the lobes retreats faster. This implies that the photosphere area that an equatorial observer sees decreases faster than that of a polar observer. 

The outcome is a rapid, and even abrupt, drop in the lightcurve for an equatorial observer, as \cite{KaplanSoker2020b} noticed with their toy model. We next demonstrate this behaviour of the lightcurve with our more sophisticated model of bipolar explosions.

\section{The lightcurves} 
\label{sec:Lightcurve}

\subsection{Assumption and scheme} 
\label{subsec:setup}

Our first assumption is that the ejecta and CSM radiates as a black body with a uniform temperature. To calculate the luminosity we need therefore to find the effective photosphere area that an observer sees. Since the morphology is, up to instabilities, axi-symmetric, we only vary the angle of the observer from the symmetry axis $\theta$ (i.e., $\theta=0$ for a polar observer and $\theta=90^\circ$ for an equatorial observer). 

For a given observation angle $\theta$ we follow many lines parallel to the line of sight that cover the entire nebula (ejecta + CSM) and find the point in the nebula along each line  where the optical depth from infinity is 
\begin{equation}
\tau = \int^{\infty}_{R_{ph}} \kappa \rho_{\rm H} dl = 2/3,
\label{eq:tauph}
\end{equation}
where $dl$ is a length element along the line of sight. We take the density $\rho_{\rm H}(x,y,z)$ from the hydrodynamical simulations (section \ref{sec:Bipolar}), and we assume a constant opacity $\kappa$. 
To these we supply two conditions. First we take only points that are in the half space closer to the observer. Namely, we take a plane perpendicular to the line of sight and through the explosion site, and consider only points with $\tau=2/3$ that are between this plane and the observer. Points in the other half of space will radiate away from the observer. Secondly, we assume that cool areas, as the hydrodynamical results give, will further cool and will not contribute to the luminosity in the visible and near IR. Therefore, in calculating the photosphere area we consider only photospheric points with a temperature of 
\begin{equation}
T_{\rm H}>T_{\rm H,c}=15,000 \K, 
\label{eq:ConditionThc}
\end{equation}
where the temperature is from the hydrodynamical simulations.  The value of $T_{\rm H,c}$ is a parameter of our lightcurve calculations that we must use because we do not include radiative transfer. Changing the value of $T_{\rm H,c}$ within the reasonable range of  $T_{\rm H,c} \la 20,000 \K$ changes the lightcurve quantitatively in a small manner, but not its qualitative behavior. The later is the goal of our present study. 

The above calculation gives us the cross section of the photosphere on the plane of the sky, i.e., the effective photospheric area $A_{\rm ef,ph}$ based on the hydrodynamical simulations and our assumptions.
To estimate the total explosion luminosity we assume that because of radiative cooling the photosphere temperature is much lower than what the hydrodynamical simulations give, and it is as observed in many CCSNe. We take $T_{\rm ef,ph}=8000 \K$, but note it might be higher or a little lower and non-uniform.
The luminosity is then 
\begin{equation}
L=4 A_{\rm ef,ph} \sigma T^4_{\rm ef,ph},
\label{eq:Luminosity}
\end{equation} 
where $\sigma$ is Stefan–Boltzmann constant. In this study we adopt 
$T_{\rm ef,ph}=8000 \K$. Later we compare to observations and calculate the absolute r-band magnitude as $m_r=4.64-2.5\log(L/L_\odot)$. 
 
{{{{ We justify the values of the two temperature parameters of $T_{\rm ef,ph}=8000 \K$ and $T_{\rm H,c}=15,000 \K$.  
Observations show that the effective temperature is not constant with time. For example, in SN 2018don that we study in section \ref{subsec:2018don}, \cite{Lunnanetal2020} infer the effective temperatures at three times to be $7500 \K$, $8000 \K$ and $6500 \K$, when they include host galaxy extinction. In the other three superluminous CCSNe that \cite{Lunnanetal2020} study the temperature decreases with time from values of $>10,000 \K$ to $\simeq 6000 \K$ weeks after explosion. As we do not include radiative transfer in this study we use a constant effective temperature. The value of $T_{\rm ef,ph}=8000 \K$ reasonably represents the type of systems we study here, and in particular SN 2018don. However, we note that generally the effective temperature decreases over time. The highest effective temperature that \cite{Lunnanetal2020} infer is $\simeq 15,000 \K$ in the first point of SN 1018bym. It is therefore reasonable to assume that the explosion itself heats the ejecta to temperatures of $\ga 15,000 \K$ in these superluminous CCSNe. Ejecta parts that start with lower temperatures cool to lower temperatures and do not contribute much to the emission. However, since we have no radiative transfer nor radiative cooling, we simply assume that the parts with temperatures of $T<T_{\rm H,c}=15,000 \K$ as we derive directly from the hydrodynamical simulations do not contribute to the lightcurve. These two values of the parameters are approximate. As we noted already, we could as well use higher values of up to $T_{\rm H,c} \simeq 20,000 \K$.  }}}}

We note that due to numerical limitations in simulating the very inner ejecta regions our results at late times are less accurate.  

\subsection{Results} 
\label{subsec:REsultsLightCurve}

In presenting the lightcurves we recall that we set $t=0$ when we start to shape the bipolar CSM inside  which the jet-driven explosion occurs. The jet-driven explosion itself takes place during the time period  $90 \days <t<93 \days$. Namely, the explosion time is $t=90 \days$. 

Before we present our results we point out again the limitations of our simple scheme to calculate the lightcurve. (1) One limitation of the hydrodynamical simulations is that we could not follow the very inner regions of the ejecta. As a result of that the lightcurves at late times are less accurate and can even be qualitatively wrong, e.g., a large increase of luminosity with time at late times.  (2) Another limitation of our lightcurve calculation is that we do not follow radiative transfer. As a result of that we do not have the correct temperature {{{{ nor the ionisation fraction }}}} in the outer regions and cannot calculate the opacity at each point. {{{{ The opacity is basically scatter on free electrons with various degrees of ionisation. }}}}  We therefore take a uniform opacity in the grid and constant in time. (3) Also, in calculating the relevant area of the photosphere, $A_{\rm ef,ph}$, we avoid low temperature regions with $T<T_{\rm H,c}$ (equation \ref{eq:ConditionThc}), where $T_{\rm H,c}$ is a parameter of our calculations. (4) We take a uniform photosphere temperature $T_{\rm ef,ph}$ that does not evolve with time. It reality it does evolve with time. 
 
\begin{figure}[htb!]
\includegraphics[trim=0.2cm 0.0cm 0.0cm 1.4cm ,clip, scale=0.33]{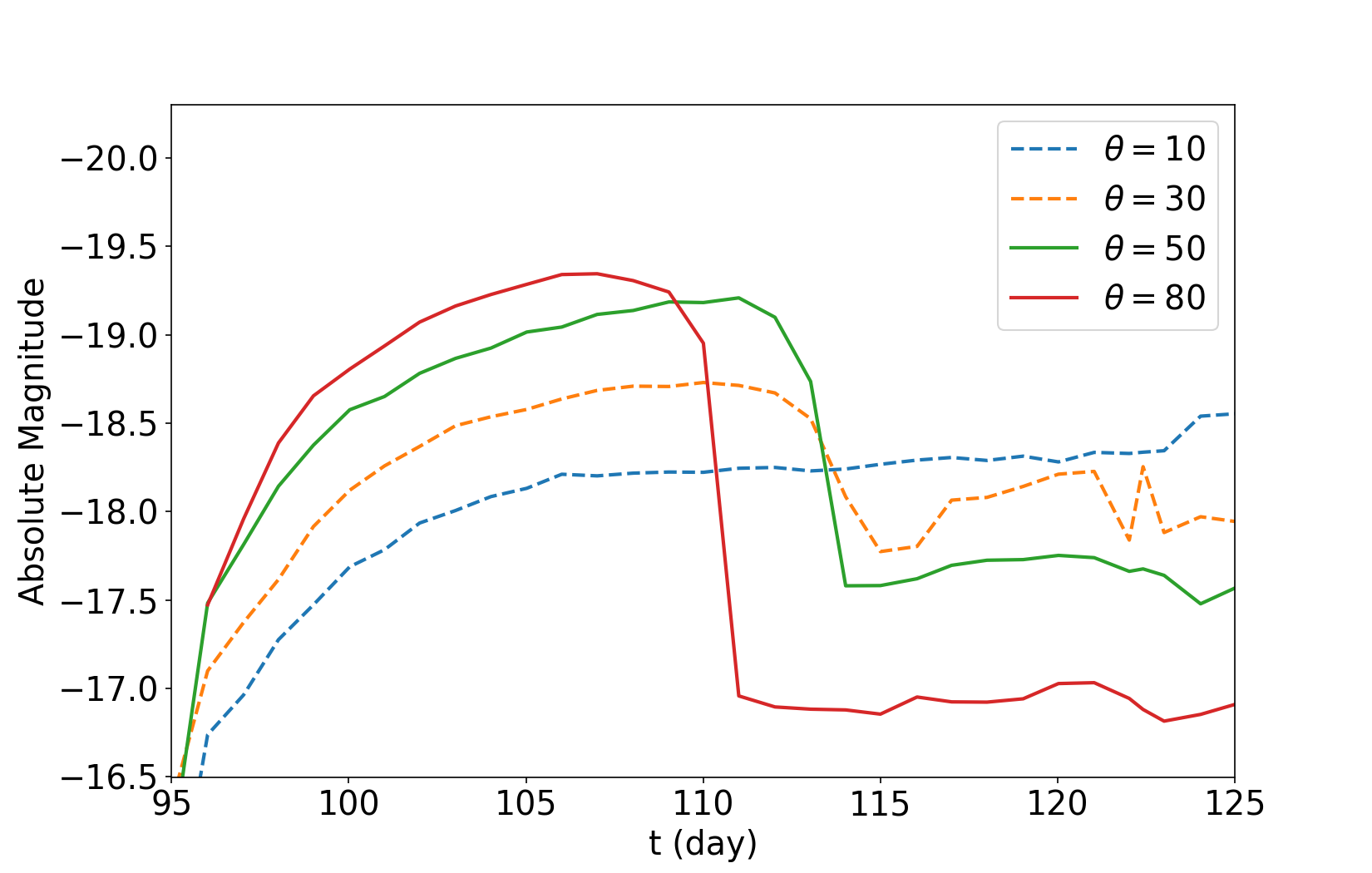}
\includegraphics[trim=0.2cm 0.0cm 0.0cm 1.4cm ,clip, scale=0.33]{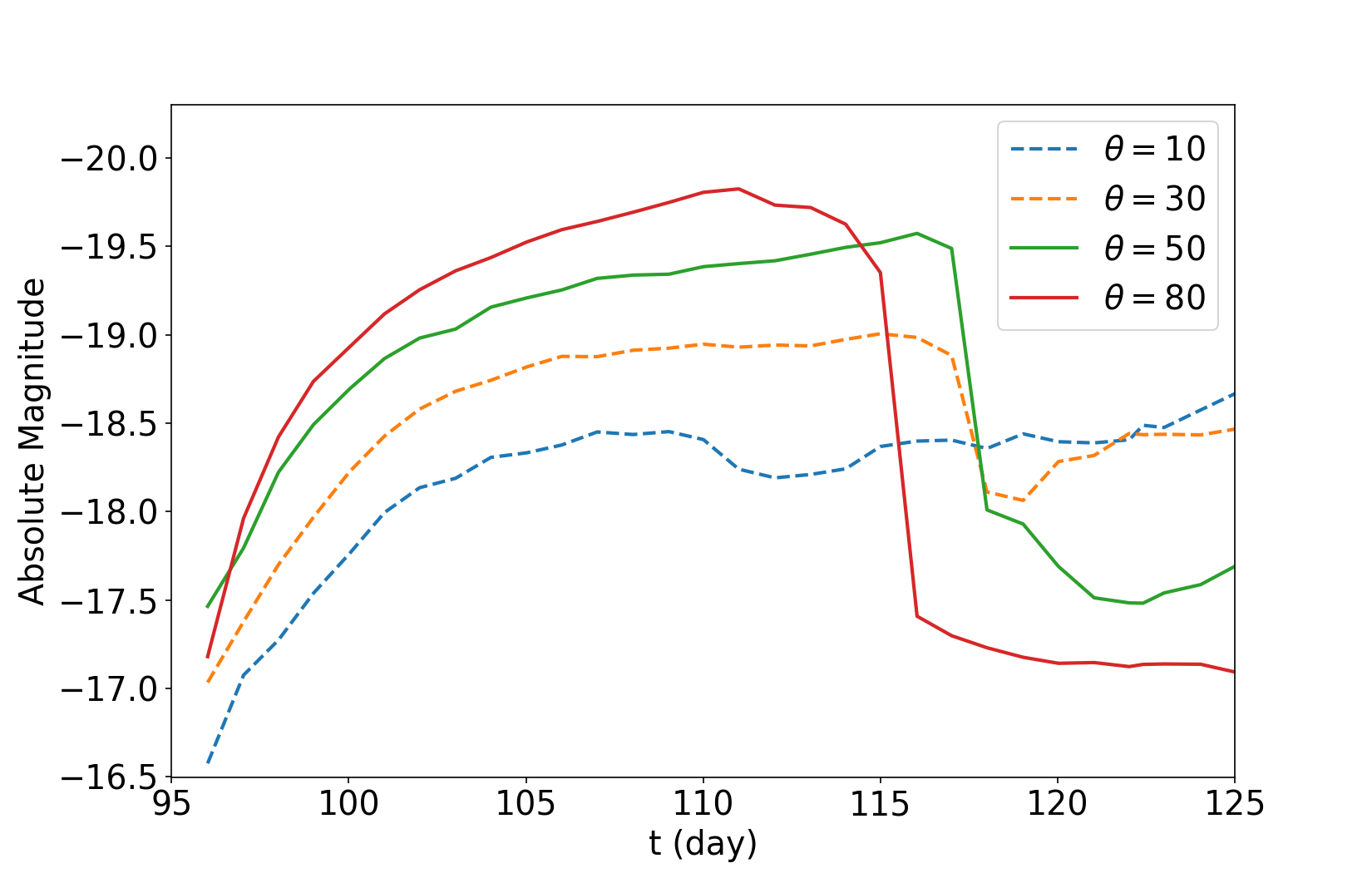}
\includegraphics[trim=0.2cm 0.0cm 0.0cm 1.4cm ,clip, scale=0.33]{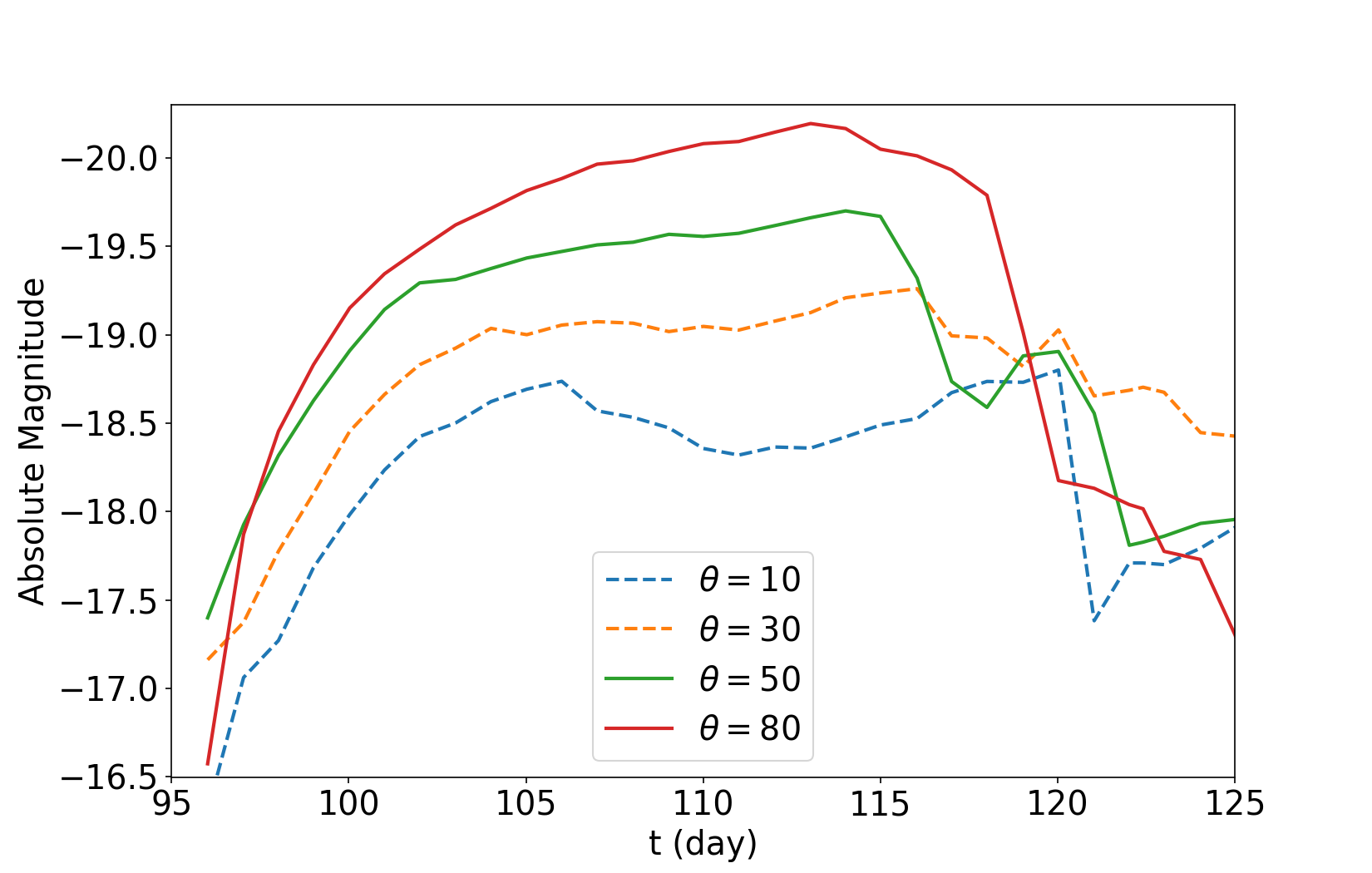}
%
	\caption{Our calculated lightcurves for the case EXP1 for observers at four different angles relative to the symmetry (polar) axis. The upper, middle, and lower panels are for opacities of $\kappa=0.06$, $0.1$, and $0.2  \cm^2 \g^{-1}$, respectively, that we use in calculating the photosphere (equation \ref{eq:tauph}). We note the rapid decline in the lightcurve for observers that are not close to the polar axis. }
\label{Fig:EXP1opacities}
\end{figure}

In Fig. \ref{Fig:EXP1opacities} we present the lightcurves (r-band magnitude) of case EXP1 for observers at four angles (measured from the polar direction), and for three opacities of $\kappa=0.06$, $0.1$, and $0.2 \cm^2 \g^{-1}$, from top to bottom panel, respectively.  As expected, the higher the opacity is the longer the broad luminosity peak is. Namely, with increasing opacity the abrupt drop in the lightcurve occurs later. 

\begin{figure}[htb!]
	\includegraphics[trim=0.2cm 0.0cm 0.0cm 1.4cm ,clip, scale=0.33]{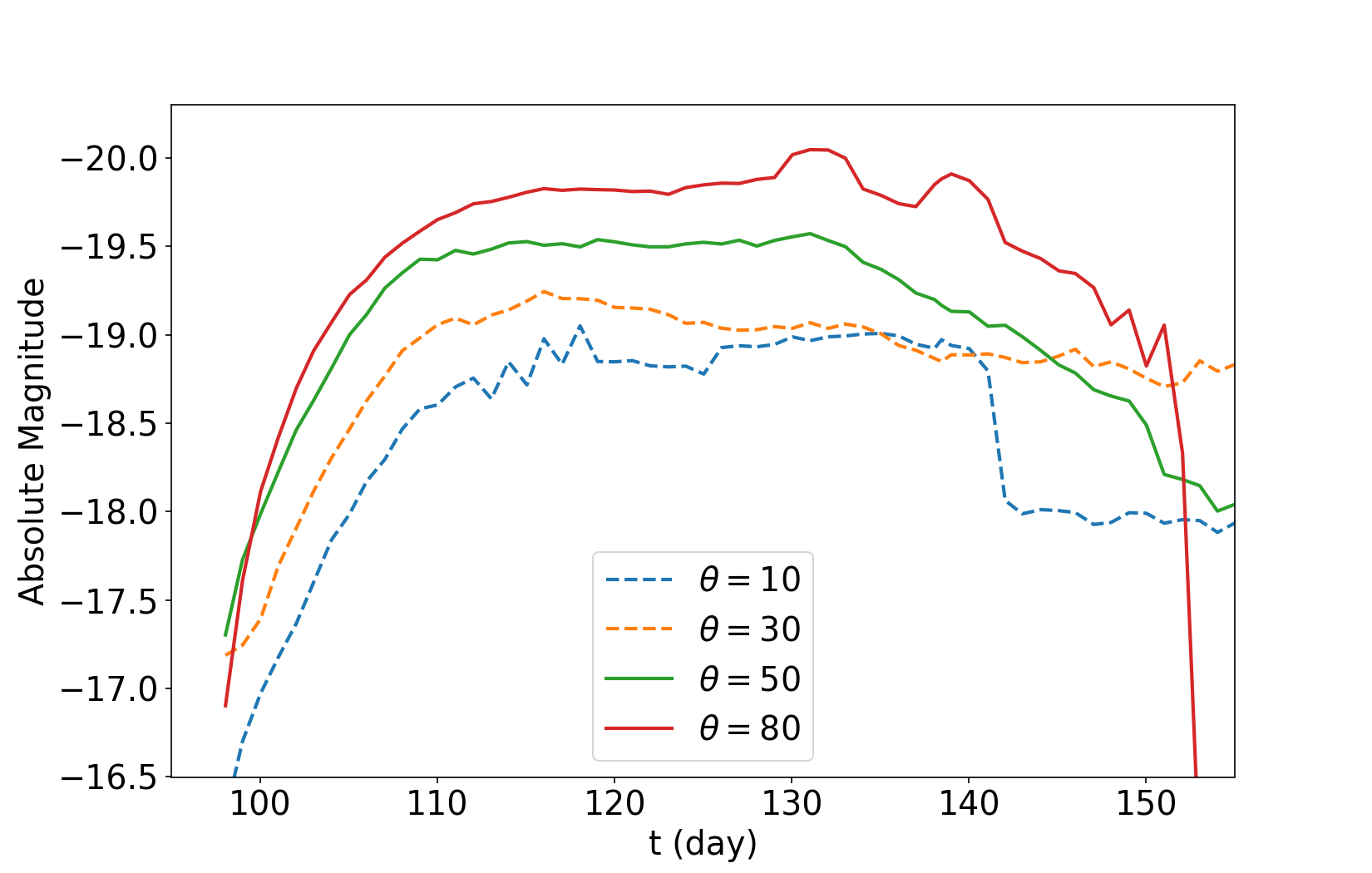}
    \includegraphics[trim=0.2cm 0.0cm 0.0cm 1.4cm ,clip, scale=0.33]{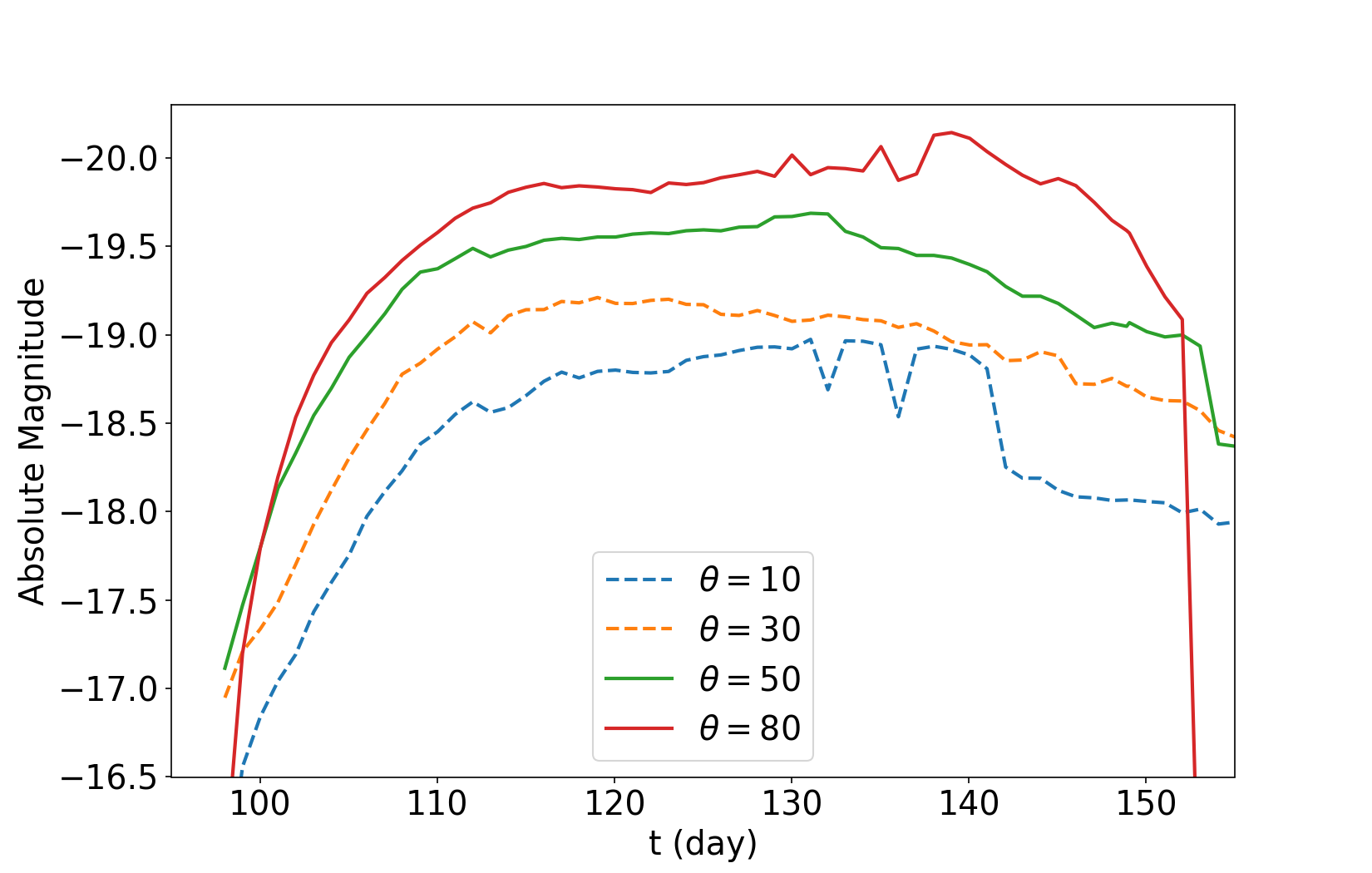}
	\includegraphics[trim=0.2cm 0.0cm 0.0cm 1.4cm ,clip, scale=0.33]{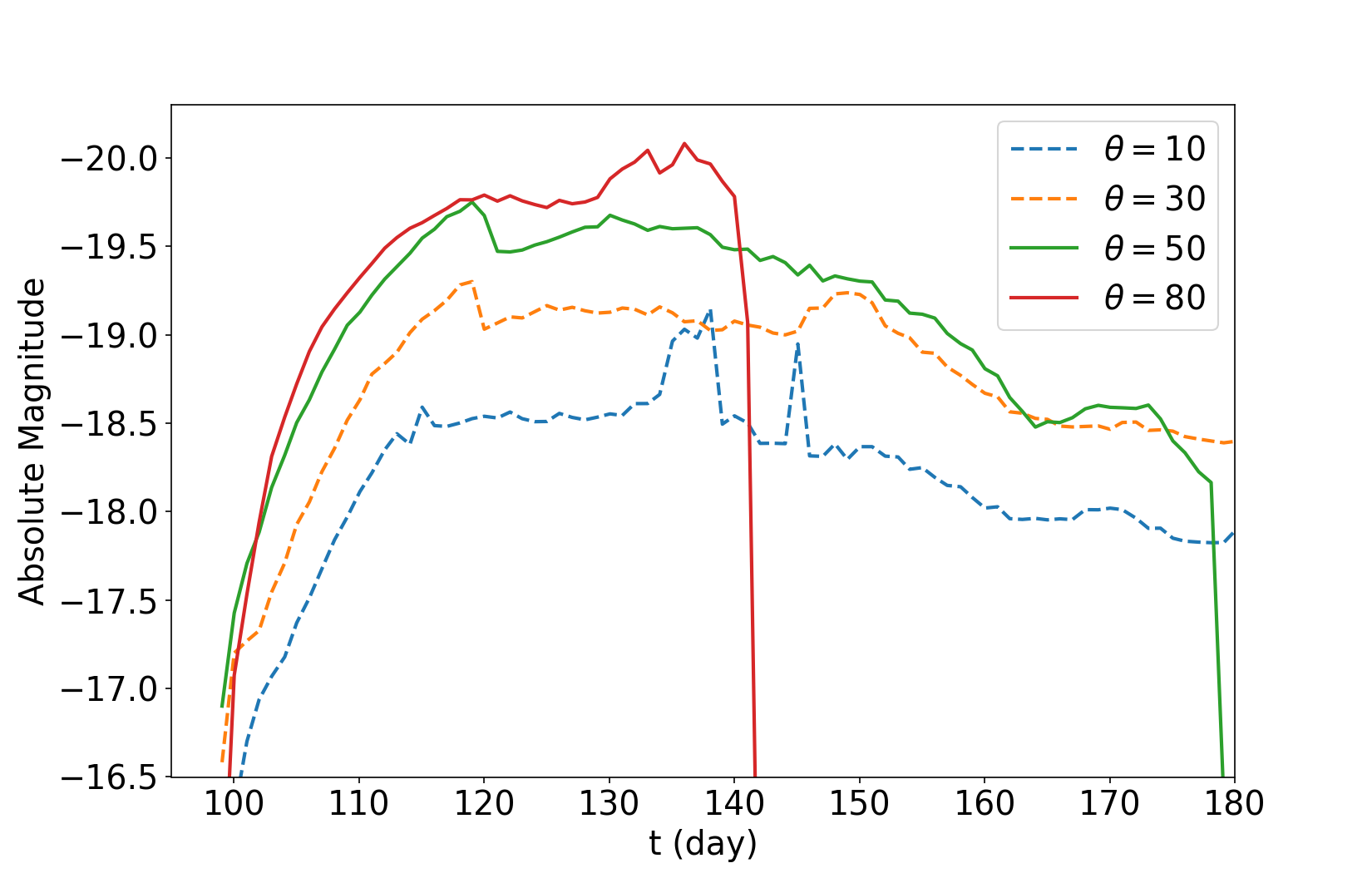}
	\caption{ The lightcurves for the cases EXP2, EXP3, and EXP4, all for an opacity of $\kappa=0.1  \cm^2 \g^{-1}$. }
	\label{Fig:MassiveCases}
\end{figure}

Another way to extend the lightcurve is to increase the mass of the CSM and/or reduce its velocity. In Fig. \ref{Fig:MassiveCases} we present the lightcurves of cases EXP2, EXP3, and EXP4 which have larger masses and EXP3 and EXP4 have also lower velocities (Table \ref{Table:cases}). These lightcurves show that a more massive CSM can prolong the wide peak up to $\simeq 80 \days$ for the parameters we use. 

Our main result is that the lightcurve for an observer that is not too close to the polar direction might have an abrupt drop. 
This is similar to the results of \cite{KaplanSoker2020b} who built a bipolar explosion toy model. We went beyond the toy model by conducting hydrodynamical simulations of jet-driven supernova explosion into a bipolar CSM (section \ref{sec:Bipolar}) and used the bipolar ejecta-CSM regions to calculate the luminosity as we explained in section \ref{subsec:setup}. Despite the limitation of our scheme, the qualitative results are robust. Exact fittings of specific cases will require more accurate calculations, in particular the inclusion of radiative transfer and a more systematic search of the (large) parameter space. Nonetheless, as we show next we can qualitatively explain the lightcurve of SN 2018don.   
 
\subsection{The case of SN 2018don} 
\label{subsec:2018don}
Below we compare our results with the lightcurve of  SN 2018don \citep{Lunnanetal2020} as \cite{KaplanSoker2020b} did.
The most similar lightcurve shapes to the shape of the lightcurve of SN 2018don are the lightcurves at $\theta=30^\circ$, $\theta=50^\circ$ and $\theta=80^\circ$ in EXP1 with $\kappa=0.1  \cm^2 \g^{-1}$ that we present in the middle panel of Fig. \ref{Fig:EXP1opacities}. 
However, the wide peak duration in these lightcurves is much  shorter than that of SN 2018don. We can prolong the wide peak duration by increasing the CSM mass, e.g., four times longer in the lower panel of Fig. \ref{Fig:MassiveCases}. However, the shapes of the lightcurves of the cases with larger masses (the three cases that we present in Fig. \ref{Fig:MassiveCases}) are not as similar to the lightcurve of SN 2018don. To obtain the same shape as in the middle panel of Fig. \ref{Fig:EXP1opacities} (case EXP1 with $\kappa=0.1 \cm^2 \g^{-1}$) and with a longer wide peak we will have to conduct a very extensive search of the parameter space. 
For example, we will have to modify also the explosion properties, which are the same in all cases here. Namely, we will have to  change in a systematic way the properties of the exploding jets, like their opening angle, energy, and duration. We also will have to consider a CSM that might be lighter in mass and that does not contain a hydrogen-rich gas. These additions are beyond our specific goal here. In future studies that will include also radiative transfer we will scan the parameter space.  

\begin{figure}[htb!]
	\centering
	\includegraphics[width=0.5\textwidth]{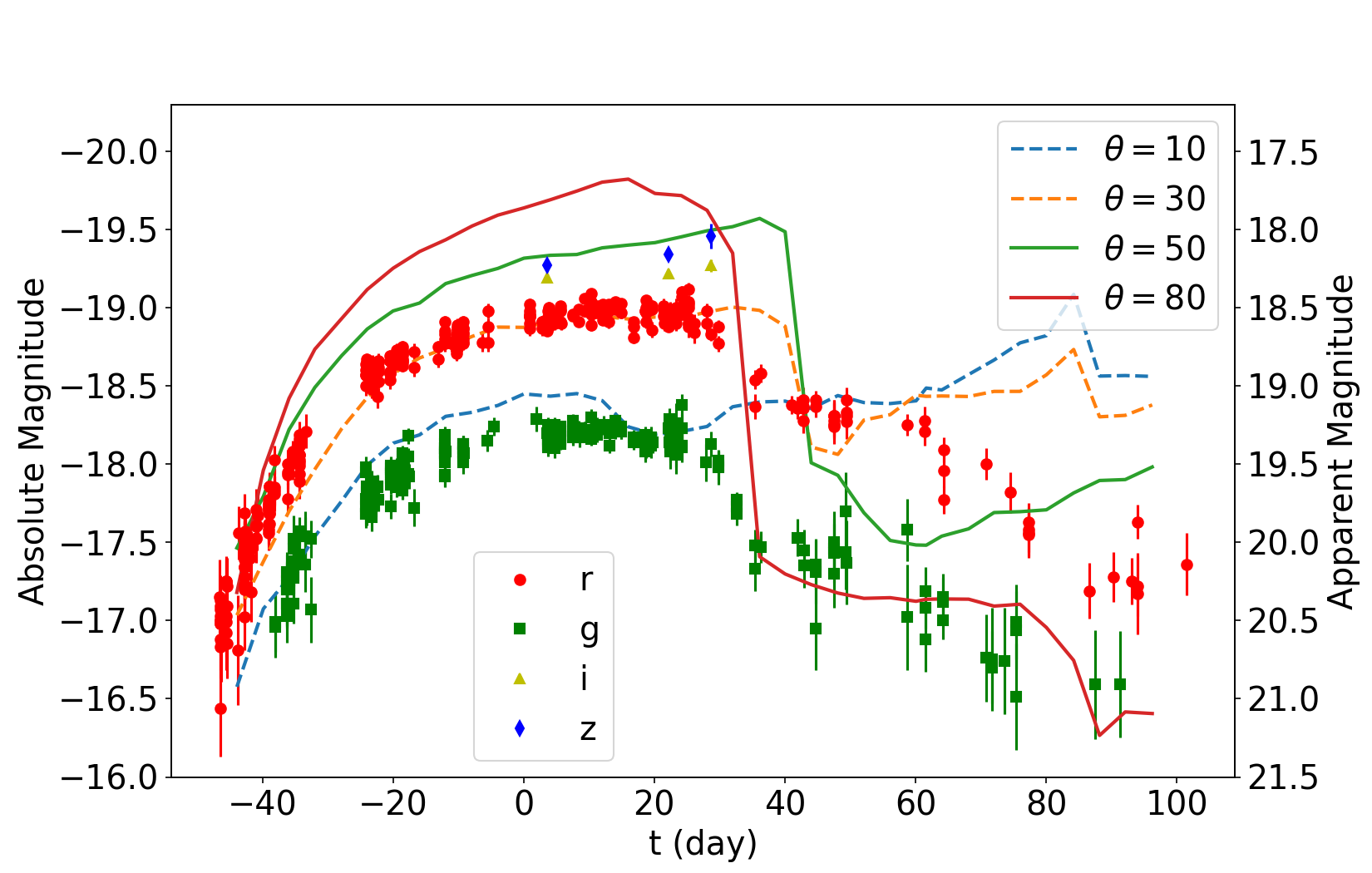}
\caption{The green and red points are the lightcurves of SN 2018don in the g-band and r-band, respectively, as
reported by \citep{Lunnanetal2020}. The four lines are based on our calculated lightcurves for the case EXP1 at four observer angles, measured from the symmetry(polar) axis as indicated in the inset, but that we stretched in time by a factor of 4. Namely, these are the four lines of the middle panel of Fig. \ref{Fig:EXP1opacities} stretched in time by a factor of 4. This figure demonstrates that we can reproduce the lightcurve shape for observers at $\theta \ga 30^\circ$, i.e., observers that are not close to the polar axis. To reproduce the correct timescale we have to take a more massive nebula as we demonstrate in Fig. \ref{Fig:MassiveCases}. 
	}
	\label{Fig:KaplanSoker}
\end{figure}

Our goal here is to present the possibility of bipolar explosions to account for lightcurves with an abrupt drop. We therefore take the lightcurves from the middle panel of Fig. \ref{Fig:EXP1opacities} and simply multiply the duration by 4 (we stretch and shift the time). We then shift the curves in time to match the zero time in the observations of SN2018don. We present our comparison in Fig. \ref{Fig:KaplanSoker}. We also note that \cite{Lunnanetal2020} estimate the maximum luminosity of SN~2018don to be $-20.1$~mag according to the extinction. 

Our conclusion from Fig. \ref{Fig:KaplanSoker} is that when we observe bipolar supernova explosions from directions that are not too close to the polar directions, the exact angle depends on the morphologies of the bipolar CSM and of the jet-driven bipolar explosion, we might have an abrupt drop in the lightcurve. 
Specific fitting of each observed lightcurve will require an extensive search of the parameter space of both the bipolar CSM, like its composition (adding hydrogen-poor gas) and other morphologies, and that of the jet-driven bipolar explosion, like other explosion energies and other jets' opening angles. 

\section{Summary} 
\label{sec:Summary}

We conducted hydrodynamical simulations of bipolar CCSN explosions and used a simple scheme to calculate the expected lightcurves. The explosion is applicable mainly (but not only) to stripped-envelope CCSNe according to the following scenario (section \ref{sec:Bipolar}). 

A companion, main sequence, neutron star or a black hole, interacts with a red supergiant star and enters a CEE with the red supergiant. The initial interaction leads to a several-years long enhanced mass loss rate by the red supergiant to all directions. In addition, the CEE that lasts weeks to months leads to an even higher mass loss rate in the equatorial plane, and at the onset of the CEE, and/or during the CEE, and/or at the exit from the CEE, the companion launches relatively weak jets that shape the CSM to possess a bipolar structure. The companion spirals-in to close distances to the core, or even merges with the core, and therefore it spins-up the core. Shortly after that the core of the red supergiant explodes. Because of the rapid pre-collapse rotation the explosion is driven by jets that have more or less a constant axis. 

We conducted four different cases of bipolar CSM hydrodynamical simulations by varying the properties of the spherical wind and/or those of the shaping jets (Table \ref{Table:cases}; section \ref{subsubsec:Wind}). We present the formation of the bipolar CSM in the case EXP1 in Fig. \ref{Fig:time_zero}. The time $t=0$ corresponds to the time when we start launching the shaping jets into the dense spherical wind (shell) to form the bipolar CSM. 

We then set the explosion by launching two energetic jets for three days starting at $t=9 0~{\rm day}$ (section \ref{subsubsec:ExplosionJets}). We launch both the shaping jets and the exploding jets (in each phase two opposite jets) along the $z$-axis. We present the evolution of the ejecta-CSM interaction for case EXP1 at four times in Figs.  
\ref{Fig:dens}-\ref{Fig:temp}.  

From the bipolar structure we calculate the projected area of the photosphere on the plane of the sky, $A_{\rm ef,ph}$, for an observer at an angle $\theta$ to the polar direction by applying conditions (\ref{eq:tauph}) and (\ref{eq:ConditionThc}), as we explain in section \ref{subsec:setup}. We then calculate the luminosity and r-band magnitude as function of time by equation (\ref{eq:Luminosity}). We present the different lightcurves in Figs. \ref{Fig:EXP1opacities} and \ref{Fig:MassiveCases}.

Our new results confirm the results of \cite{KaplanSoker2020b} that the lightcurve of an equatorial observer of a bipolar CCSN might have a rapid, and even an abrupt, drop. \cite{KaplanSoker2020b} used a bipolar toy model. We here calculated the lightcurves from bipolar explosions that we built with 3D hydrodynamical simulations that we base on a CEE scenario. In that respect our calculation are more realistic. However, since we did not include radiative transfer we had to assume a uniform and constant opacity in equation (\ref{eq:tauph}), a minimum temperature $T_{\rm H,c}$ (equation \ref{eq:ConditionThc}) from the hydrodynamical simulations for a photosphere area element to be included in the effective projected area $A_{\rm ef,ph}$, and a photospheric temperature $T_{\rm ef,ph}$ in the calculation of the luminosity (equation \ref{eq:Luminosity}).
All the values that we assume are within a very reasonable range of possible parameters of CCSNe. Due to numerical limitations we could not follow in details the inner regions of the ejecta, and for that our lightcurves at late times are not accurate. 
 
Like \cite{KaplanSoker2020b} we try to fit the lightcurve of the hydrogen-poor SN 2018don. The lightcurves shapes that best fit the lightcurves are those of case EXP1 (Fig. \ref{Fig:EXP1opacities}). However, these lightcurves evolve too rapidly. We showed that by increasing the mass in the CSM and/or reducing its velocity we can prolong the lightcurve (Fig. \ref{Fig:MassiveCases}). Because of the numerical difficulties we could not obtain lightcurves with similar shapes to that of SN2018do. For that, we simply took the lightcurve from the middle panel of Fig. \ref{Fig:EXP1opacities} and stretched the time by a factor of 4. In order to fit the lightcurve of SN2018 we will have to make an extensive search of the parameter space, including changing the properties of the exploding jets and compositions of the CSM and ejecta, something we did not do here. 

Let us place our results within a broader scope. {{{{ Some recent studies argue }}}} that modelling lightcurves with the neutrino-driven explosion mechanism and then adding a magnetar and/or ejecta-CSM interaction cannot explain a large fraction of LSNe and SLSNe (for a review of this see \citealt{Soker2022Rev}). 
{{{{ \cite{SokerGilkis2017a} find that in about half of the SLSNe lightcurves  that \cite{Nicholletal2017b} fit with a magnetar the CCSN explosion  energy is $E_{\rm SN}  > 2 \times 10^{51} \erg$. This is more than what the delayed neutrino mechanism can account for, and therefore \cite{SokerGilkis2017a} conclude that jets exploded these SLSNe. The magnetohydrodynamic simulations by \cite{Reichertetal2022} support this claim for jet-powering. Similarly, \cite{Soker2022LSNe} examines the  the lightcurves fitting with magnetars of 40 LSNe by \cite{Gomezetal2022}, and finds inconsistency in 18 LSNe out of 40. 
Further to these,  \cite{Soker2022Rev} examined the fitting to the lightcurves of 70 hydrogen-poor SLSNe that \cite{Chenetal2022} perform with magnetars and/or ejecta-CSM. \cite{Soker2022Rev} argues that even the slower magnetars in their fitting most likely launched jets at explosion. In 7 out of 14 SLSNe for which \cite{Chenetal2022} apply ejecta-CSM interaction to power the lightcurve, \cite{Soker2022Rev} finds the kinetic energy of the ejecta to be larger than what the delayed neutrino explosion mechanism can supply. With these claims for difficulties of the magnetar and ejecta-CSM fittings to account for about half of SLSNe, and in particular for the neutrino delayed mechanism to account for the required large explosion energy in many cases, we accept the possibility that jets explode a large fraction of SLSNe. }}}}
Our study strengthens the claim that jet-driven explosions can account for some properties of LSNe and SLSNe.  


\section*{Acknowledgments}

{{{{ We thank an anonymous referee for the good suggestions and comments. }}}}
This research was supported by the Amnon Pazy Research Foundation.
AM acknowledges support from the R\&D Authority in Ariel University. 
Numerical simulations presented in this work were per-formed on the Hive computer cluster at the University of Haifa.

\section*{Data availability}

The data underlying this article will be shared on reasonable request to the corresponding author.  


\end{document}